\documentclass[12pt]{article}

\usepackage[height=8.85in,width=6.45in]{geometry}
\usepackage{fancybox}
\usepackage{amsmath,amssymb}
\usepackage{comment}
\usepackage{hyperref}

\usepackage{xcolor}

\newcommand{\cS}{\mathcal{S}}
\newcommand{\cQ}{\mathcal{Q}}
\newcommand{\tcS}{\tilde{\mathcal{S}}}
\newcommand{\tcQ}{\tilde{\mathcal{Q}}}
\newcommand{\cR}{\mathcal{R}}
\newcommand{\cM}{\mathcal{M}}
\newcommand{\cO}{\mathcal{O}}

\newcommand{\cT}{\mathcal{T}}
\newcommand{\cK}{\mathcal{K}}
\newcommand{\cP}{\mathcal{P}}
\newcommand{\cJ}{\mathcal{J}}
\newcommand{\so}{\mathfrak{so}}

\newcommand{\cH}{\mathcal{H}}

\usepackage{times}
\usepackage{courier}
\usepackage{mathtools}
\numberwithin{equation}{section}

\let\bar\overline

\def\Nequals#1{$\mathcal{N}{=}\,#1$}

\newcommand{\bC}{\mathbb{C}}
\newcommand{\bR}{\mathbb{R}}
\newcommand{\bZ}{\mathbb{Z}}

\newcommand{\cD}{\mathcal{D}}

\newcommand{\SL}{\mathrm{SL}}
\newcommand{\SU}{\mathrm{SU}}
\newcommand{\U}{\mathrm{U}}
\newcommand{\SO}{\mathrm{SO}}

\newcommand{\su}{\mathfrak{su}}

\begin{document}

\begin{titlepage}

\begin{flushright}
IPMU-16-0009\\
UT-16-03\\
YITP-16-8
\end{flushright}

\vskip 2cm

\begin{center}

{\Large On 4d rank-one \Nequals3  superconformal field theories }

\vskip 1cm
Takahiro Nishinaka$^1$ and  Yuji Tachikawa$^{2,3}$
\vskip 1cm

\begin{tabular}{ll}
1. & Yukawa Institute for Theoretical Physics,\\
& Kyoto University, Kyoto 606-8502, Japan\\
 2. & Department of Physics, Faculty of Science, \\
& University of Tokyo,  Bunkyo-ku, Tokyo 113-0033, Japan\\
3.  & Kavli Institute for the Physics and Mathematics of the Universe, \\
& University of Tokyo,  Kashiwa, Chiba 277-8583, Japan\\
\end{tabular}

\vskip 1cm

\textbf{Abstract}

\end{center}

\vskip1cm

\noindent
We study the properties of 4d \Nequals3 superconformal field theories whose rank is one, i.e.~those that reduce to a single vector multiplet on their moduli space of vacua. 
We find that the moduli space can only be of the form $\bC^3/\bZ_{\ell}$ for $\ell{=}1,2,3,4,6$, and
that the supersymmetry automatically enhances to \Nequals4 for $\ell{=}1,2$.
In addition, we determine the central charges $a$ and $c$ in terms of $\ell$, 
and construct the associated 2d chiral algebras, which turn out to be exotic \Nequals2 supersymmetric W-algebras. 

\end{titlepage}

\setcounter{tocdepth}{2}
\tableofcontents

\section{Introduction and summary}
Four-dimensional non-gravitational theories with \Nequals3 supersymmetry have been mostly neglected in the literature, due to the well-known fact that any \Nequals3 supersymmetric Lagrangian automatically possesses \Nequals4 supersymmetry. 
The developments in the last several years on the supersymmetric dynamics tell us, however, that there are many `non-Lagrangian' theories, i.e.~strongly-coupled field theories which do not have obvious Lagrangian descriptions.

Therefore there can be non-Lagrangian \Nequals3 theories, some of whose general properties were first discussed in a paper by Aharony and Evtikhiev \cite{Aharony:2015oyb} from early December 2015. 
Later in the same month, Garc\'\i a-Etxebarria and Regalado made a striking discovery \cite{Garcia-Etxebarria:2015wns} that indeed such \Nequals3 theories appear on D3-branes probing a generalized form of orientifolds in F-theory.\footnote{Related holographic constructions of \Nequals3 systems were already discussed in a paper \cite{Ferrara:1998zt} from 1998, although no concrete models were identified there.  The authors thank T.~Nishioka for bringing this reference \cite{Ferrara:1998zt} to their attention. Also see a recent paper \cite{Beck:2016lwk} discussing \Nequals3 holographic duals in (massive) type IIA and type IIB setups.}

The aim of this note is to initiate the  analysis of  such concrete \Nequals3 theories in a purely field-theoretical manner. 
We mainly restrict attention to rank-1 theories, where the rank is defined as the dimension of the Coulomb branch of the theory considered as an \Nequals2 theory.
We will find 
\begin{itemize}
\item that the moduli space of supersymmetric vacua can only be of the form $\bC^3/\bZ_\ell $ for $\ell=1,2,3,4,6$,
\item that the supersymmetry is guaranteed to enhance to \Nequals4 for $\ell=1,2$, and therefore only $\ell=3,4,6$ are allowed in the case of the genuine \Nequals3 theories,
\item and that the central charges are given by $a=c=(2\ell-1)/4$.
\end{itemize}

In addition we construct the 2d chiral algebras associated in the sense of \cite{Beem:2013sza} to these rank-1 \Nequals3 theories. We will find the following:
\begin{itemize}
\item The 2d chiral algebra contains the \Nequals2 super Virasoro subalgebra and a pair of bosonic chiral primary and antichiral primary with dimension $\ell/2$, as a consequence of 
 the unitarity bounds and the operator product expansions of the 4d \Nequals3 superconformal algebra.
\item The Jacobi identities of these operators close only for a finite number of central charges, including $c_{2d}=-3(2\ell-1)$ as predicted from the construction of \cite{Beem:2013sza}. Furthermore,  the null relation correctly encodes the structure of the moduli space of supersymmetric vacua at this value of the central charge. 
\end{itemize}
Further studies of these chiral algebras will uncover the spectrum of BPS local operators in rank-1 \Nequals3 superconformal field theories (SCFTs), along the lines of \cite{Beem:2013sza, Beem:2014rza, Lemos:2014lua, Buican:2015ina, Buican:2015hsa, Cordova:2015nma, Buican:2015tda, Song:2015wta}.

All the findings in this note are consistent  with, but do not prove, the existence of genuine \Nequals3 theories with $\ell=3,4,6$. 
We also note that the findings do not preclude the existence of multiple distinct \Nequals3 theories with the same value of $\ell$, although the data we compute in this note do not distinguish them.

The rest of the note is organized as follows: 
in Sec.~\ref{moduli}, we study basic properties of \Nequals3 rank-1 theories. We see that the moduli space of supersymmetric vacua is necessarily of the form $\bC^3/\bZ_{\ell}$ for $\ell=1,2,3,4,6$, with an automatic enhancement to \Nequals4 when $\ell=1,2$. 
In Sec.~\ref{unitary}, we analyze the shortening conditions and the unitarity bounds of the \Nequals3 superconformal algebras to the extent necessary for us, and a few general properties of the associated 2d chiral algebra. 
In Sec.~\ref{jacobi}, we use the results obtained so far to construct the 2d chiral algebra associated to \Nequals3 rank-1 theories for $\ell=3,4,6$.

\paragraph{Note added:} When this paper is completed, the authors learned from P. Argyres, M. Lotito, Y. L\"u and M. Martone that they have an upcoming paper \cite{ArgyresEtAlToAppear} which has a small overlap with but is largely complementary to this paper.  The authors thank them for sharing the draft in advance.

\section{Basic properties}\label{moduli}
\subsection{Allowed forms of the moduli space}
Let us start by analyzing the allowed form of the moduli space of vacua of an \Nequals3 rank-1 superconformal field theory. 
Regarding it as an \Nequals2 theory, its Coulomb branch should be a one-dimensional scale-invariant Seiberg-Witten geometry.
Its classification is well-known: one just needs to go through Kodaira's list of singularities of elliptic fibrations and keep only the ones where the modulus of the elliptic fiber is constant.
The resulting list is reproduced in Table~\ref{scale-inv-kodaira}.
In particular, the scaling dimension of the Coulomb branch operator $u$ is fixed to be one of the eight possible values listed there.

\begin{table}
\[
\begin{array}{c|ccccccccccc}
& I_0 & II & III & IV & I_0^* & IV^* & III^* & II^*\\
\hline
\Delta(u) &1 & 6/5 & 4/3 & 3/2 & 2 & 3 & 4 & 6 \\
\tau & \text{arb.} & \omega & i& \omega & \text{arb.} & \omega & i & \omega\\
g & \begin{psmallmatrix}
1 & 0 \\
0 & 1
\end{psmallmatrix} 
& \begin{psmallmatrix}
1 & 1 \\
-1 & 0
\end{psmallmatrix}
& \begin{psmallmatrix}
0 & 1 \\
-1 &0 
\end{psmallmatrix} 
& \begin{psmallmatrix}
0 & 1 \\
-1 & -1 
\end{psmallmatrix}
& \begin{psmallmatrix}
-1 & 0 \\
0 & -1 
\end{psmallmatrix}
& \begin{psmallmatrix}
-1 & -1 \\
1 & 0
\end{psmallmatrix}
& \begin{psmallmatrix}
0 & -1 \\
1 & 0 
\end{psmallmatrix} 
& \begin{psmallmatrix}
0 & -1 \\
1 & 1
\end{psmallmatrix}
\end{array}
\]
\caption{The list of scale invariant rank-1 Seiberg-Witten geometries. The first line shows the name given by Kodaira; $\Delta(u)$ is the scaling dimension of the Coulomb branch operator $u$; $\tau$ is the complexified coupling at the generic points on the Coulomb branch; and $g$ is the $\SL(2,\bZ)$ monodromy around the origin.  On the row for $\tau$, $\omega$ is a third root of unity, and \text{arb.} means that $\tau$ is arbitrary. \label{scale-inv-kodaira}}
\end{table}

The \Nequals3 supersymmetry relates the Higgs branch and the Coulomb branch of the theory regarded as an \Nequals2 theory. 
The Higgs branch at the origin $u=0$ of the Coulomb branch is then a hyperk\"ahler cone of quaternionic dimension one. 
Such a one-dimensional cone is necessarily an asymptotically locally Euclidean space of the form $\bC^2/\Gamma$ where $\Gamma$ is a discrete subgroup of $\SU(2)$. 
As an \Nequals3 supersymmetric theory necessarily has a $\U(1)$ flavor symmetry as seen as an \Nequals2 theory, the space $\bC^2/\Gamma$ should have a $\U(1)$ hyperk\"ahler isometry.
This restricts $\Gamma$ to be of the form $\bZ_\ell $.
Let $(z_+,z_-)$ be the coordinates of $\bC^2$ before the quotient.
Then, as an \Nequals1 theory, the Higgs branch are parameterized by three chiral operators $W_+=z_+^\ell $, $W_-=z_-^\ell$ and $J=z_+z_-$ satisfying \begin{equation}
W^+ W^- \propto J^\ell.\label{higgs-branch-relation}
\end{equation} Here, $W^\pm$ has dimension $\ell$ and $\U(1)$ charge $\pm\ell$,  and $J$ is the moment map of the $\U(1)$ symmetry.

The \Nequals3 symmetry rotates the \Nequals2 Coulomb branch to the \Nequals2 Higgs branch, and therefore relates the operator $W^\pm$ and $u$, as we will see this in more detail in Sec.~\ref{subsubsec:unitarity}.
This means that the integer $\ell$ should also be an number allowed as $\Delta(u)$. 
We conclude that $\ell=1,2,3,4,6$. 
Combining the information on the Coulomb branch and the Higgs branch, we see that the full moduli space of supersymmetric vacua should be of the form $\bC^3/\bZ_\ell $, where $\ell$ is one from the list above.

That the moduli space of \Nequals3 theory is locally flat is known, see e.g.~\cite{CecottiTextbook}.
Let us check that the quotient by $\bZ_\ell $ preserves \Nequals3 supersymmetry, at least away from the origin.\footnote{The analysis of the supercharges here is completely the same as the one given in Garc\'\i a-Etxebarria and Regalado \cite{Garcia-Etxebarria:2015wns} done in F-theory. The point here is that it can be phrased in a completely field-theoretical manner.} 
Note that away from the origin, the moduli space is smooth. 
As such, the theory is locally that of a single \Nequals4 $\U(1)$ vector multiplet.
Let us denote the three chiral scalars of the vector multiplet by $(z_0,z_+,z_-)$, 
such that $u=z_0^\ell $.
The $\bZ_\ell $ action acts as \begin{equation}
(z_0,z_+,z_-) \mapsto (\gamma z_0,\gamma z_+,\gamma^{-1} z_-)\label{geom-action-of-k}
\end{equation} where $\gamma=e^{2\pi i/\ell}$.
This is accompanied by an $\SL(2,\bZ)$ duality action $g$ given in Table~\ref{scale-inv-kodaira}.

The geometric action \eqref{geom-action-of-k} is a part of the $\SU(4)$ R-symmetry of the free \Nequals4 multiplet, which determines its action of the four supercharges as \begin{equation}
(Q_1,Q_2,Q_3,Q_4)\mapsto (\gamma^{1/2} Q_1,\gamma^{1/2} Q_2,\gamma^{1/2} Q_3, \gamma^{-3/2} Q_4).
\end{equation}
The action of the duality transformation by $g$ on the supercharges can be found e.g.~in Sec.~2.2 of \cite{Kapustin:2006pk}:
\begin{equation}
(Q_1,Q_2,Q_3,Q_4)\mapsto \gamma^{-1/2}(Q_1,Q_2,Q_3,Q_4).
\end{equation}
 Combined, we see that the action on the four supercharges is given by \begin{equation}
(Q_1,Q_2,Q_3,Q_4)\mapsto (Q_1,Q_2,Q_3, \gamma^{-2} Q_4),
\end{equation} from which we conclude that all four supercharges are preserved for $\ell=1,2$ whereas only the first three supercharges are preserved for $\ell=3,4,6$.

The enhancement to \Nequals4 when $\ell=1,2$ can be understood also as follows.
When $\ell=1,2$, the hyperk\"ahler cone $\bC^2/\bZ_\ell $ has a larger hyperk\"ahler isometry $\SU(2)$
with corresponding moment map operators of dimension two.
This in turn implies that the flavor symmetry as an \Nequals2 theory is larger than $\U(1)$.
In \cite{Aharony:2015oyb} it was shown that genuine \Nequals3 theories cannot have any flavor symmetry current bigger than $\U(1)$ as \Nequals2 theory, 
meaning that the supersymmetry automatically enhances to \Nequals4 for $\ell=1,2$. 

Finally, let us determine the central charges $a$ and $c$ of these theories labeled by $\ell$.
Very generally, any \Nequals2 superconformal field theory is believed to satisfy the relation \begin{equation}
2a-c = \frac14\sum_i (2\Delta(u_i)-1 )
\end{equation} where the sum runs over the independent generators of the Coulomb branch operators.
 
This relation was originally conjectured in \cite{Argyres:2007tq} and a derivation that applies to a large subclass of \Nequals2 theories was given in \cite{Shapere:2008zf}. 
It is not perfectly clear that the assumptions used in \cite{Shapere:2008zf} is satisfied by strongly-coupled theories we are discussing here, but the authors think it is quite plausible.\footnote{%
It is known that this relation fails in  gauge theories where part of the gauge symmetry is disconnected from the identity.
For example, take \Nequals4 super Yang-Mills theory with gauge group $\U(1)$ and $\mathrm{O}(2)$.
They both have $2a-c=1/4$, but the Coulomb branch operator has dimension $1$ for the former and $2$ for the latter.
In this note, when we speak about the moduli space of vacua, we declare that we do \emph{not} impose the invariance under the disconnected part of the gauge group, or whatever that concept corresponds to in non-Lagrangian theories. 
The author expects that this relation holds under this condition.
\label{subtlety}
}
Assuming the validity of the general formula, we then have \begin{equation}
2a-c = \frac{2\ell-1}{4}.
\end{equation}

Now, in any \Nequals3 superconformal field theory, we have $a=c$, as originally shown in \cite{Aharony:2015oyb}. One way to re-derive it in our case is to go to the Higgs branch as an \Nequals2 theory. 
This process does not break $\U(1)_R$ symmetry in the \Nequals2 subalgebra,
and hence the $\U(1)_R$-gravity-gravity anomaly, which is proportional to $a-c$, is conserved. 
On the Higgs branch the theory is just \Nequals4, and therefore $a-c=0$.

From the known value of $2a-c$ above, we conclude that \begin{equation}
a=c=\frac{2\ell-1}{4}.\label{centralcharges}
\end{equation} 
As mentioned above, the derivation here is not completely watertight, but we give a rather non-trivial consistency check in the rest of the paper.

\subsection{Realizations}
So far we concluded that the moduli space of a rank-1 \Nequals3 superconformal field theory is necessarily of the form $\bC^3/\bZ_\ell $ for $\ell=1,2,3,4,6$. 
Here we give a brief survey of the known realizations of these theories. 

When $\ell=1,2$, the theory automatically has \Nequals4 supersymmetry.
For $\ell=1$, the vacuum moduli space is simply $\bC^3$ without any singularity, and therefore we can safely conclude that the only such theory is a theory of a single free $\U(1)$ vector multiplet.
For $\ell=2$, a realization is of course given by the \Nequals4 super Yang-Mills theory with gauge algebra $\mathfrak{su}(2)$. 
The gauge group can either be $\SU(2)$ or $\SO(3)$, depending on which we have two subtly different theories.\footnote{%
We declare that the \Nequals4 super Yang-Mills theory with gauge group $\mathrm{O}(2)$ belongs to the case $\ell=1$, as discussed in Footnote~\ref{subtlety}.
}

Genuine \Nequals3 theories were first constructed in \cite{Garcia-Etxebarria:2015wns} using F-theory. 
Namely, they started from the F-theory setup of the form $\bR^{1,3}\times \bC^3 \times T^2$
where the last $T^2$ describes the axiodilaton of the Type IIB theory, took the quotient $(\bC^3\times T^2)/\bZ_k$, and probed this background by $r$ D3-branes. 
In particular, we have rank one theories when $r=1$, and the moduli space of vacua is parameterized by the position of the D3-brane, that is $\bC^3/\bZ_k$.
As the torus $T^2$ can have $\bZ_k$ isometry only for $k=1,2,3,4,6$, 
we get the same classification as we saw above.

There is a caveat however:
we cannot directly identify the integer $k$ governing the F-theory background
and the integer $\ell$ governing the moduli space of the superconformal theory.
When $k=2$, there are two types of such $\bZ_2$ quotient, up to the action of the $\SL(2,\bZ)$ duality of the type IIB. One is the O3$^-$ plane and the other is the $\SL(2,\bZ)$ orbit containing O3$^+$, $\widetilde{\text{O3}}^-$, $\widetilde{\text{O3}}^+$.
Probing by one D3-brane, the former gives the \Nequals4 super Yang-Mills theory with gauge algebra $\mathfrak{so}(2)$, whereas the latter gives that with gauge algebra $\mathfrak{su}(2)$.
As discussed in Footnote~\ref{subtlety}, we declare that when we discuss the moduli space we do not gauge by the disconnected part of the gauge group, and then the former has the moduli space $\bC^3/\bZ_{\ell=1}$
whereas the latter has $\bC^3/\bZ_{\ell=2}$. 
In both cases, $\ell$ divides $k$.

As discussed in \cite{Garcia-Etxebarria:2015wns}, there are various versions of the $\bZ_k$ quotients also for $k\neq 2$ in F-theory. Depending on the version, we will have a different discrete quotient \begin{equation}
\bC^3/\bZ_\ell  \to \bC^3/\bZ_k
\end{equation} where the left hand side is the moduli space of the superconformal theory
and the right hand side is the F-theory background.
We do not yet know which version of the $\bZ_k$ quotient gives which divisor $\ell$ of $k$.

If there would be a version such that $k=\ell$ for each $\ell=3,4,6$,
 we would have an F-theoretic realization of an \Nequals3 rank-1 theory for each $\ell=3,4,6$.
This point is however not well understood and requires further study, and the details will be reported elsewhere \cite{AharonyTachikawaWorkInProgress}.
We would like to point out that, even assuming this, the F-theory construction gives \emph{a} realization; we do not yet know whether there are multiple subtly different versions of the theory for each $\ell=3,4,6$ either. 

The rank-1 \Nequals3 theories are already quite interesting even when considered as \Nequals2 theories,
since they give rise to rank-1 \Nequals2 theories in addition to the known list consisting of the old ones \cite{Argyres:1995jj,Argyres:1995xn,Minahan:1996fg,Minahan:1996cj} and the new ones \cite{Argyres:2007tq,Argyres:2010py}.\footnote{See \cite{Chacaltana:2016shw} for an even newer rank-1 \Nequals2 theory with $\SU(4)$ symmetry.}
As already discussed, the \Nequals3 theory would have $\mathfrak{u}(1)$ flavor symmetry as an \Nequals2 theory.

A systematic study of all possible rank-1 \Nequals2 superconformal field theories and their mass deformations through the construction of the Seiberg-Witten curves and differentials are being carried out by Argyres, Lotito, L\"u and Martone \cite{Argyres:2015ffa,Argyres:2015gha}.
The properties of the $\ell=3$ theory we determined above match exactly with the entry in Table~1 of \cite{Argyres:2015ffa} describing the $IV^*$ singularity with $\mathfrak{u}(1)$ symmetry.
The $\ell=4$ and the $\ell=6$ theories might similarly correspond to some of  the entries in the same Table.  
We immediately notice, however, that there are no entries of the $III^*$ and $II^*$ singularities marked there as having $\mathfrak{u}(1)$ flavor symmetry.
This does not yet preclude the existence of the $\ell=4$ and $\ell=6$ theories, since in \cite{Argyres:2015ffa,Argyres:2015gha} it was assumed that all the discrete symmetries acting on the mass parameters were considered as coming from the Weyl symmetry.
In particular, in their construction, those marked as having $\mathfrak{su}(2)$ flavor symmetry can be interpreted as having $\bZ_2 \ltimes \U(1)$ symmetry. 
This point clearly needs further study.\footnote{The authors thank P. Argyres, M. Lotito, Y. L\"u and M. Martone for instructive discussions on this point, and for sharing their upcoming paper \cite{ArgyresEtAlToAppear}.}

\section{4d $\mathcal{N}{=}3$ theories and the associated 2d chiral algebras}\label{unitary}

In this section, we work out some consequences of the 4d \Nequals3 superconformal algebras 
and state them in the \Nequals2 language. We also derive general properties of the 2d chiral algebra $\chi[\mathcal{T}]$ associated in the sense of \cite{Beem:2013sza} to an \Nequals3 superconformal theory $\mathcal{T}$.
We mainly follow the convention of \cite{Dolan:2002zh} here.

\subsection{\Nequals3 superconformal algebra and its \Nequals2 subalgebra}
\label{subsubsec:unitarity}

\begin{table}
\begin{center}
\begin{tabular}{|c|c|c|c|c|c|c|c|}
\hline
 & $j_1$& $j_2$ & $R$ & $r$ & $F$ & $\delta_1$ & $\delta_2$ \\
\hline
\hline
$\mathcal{Q}^1{}_+$ & $+ \frac{1}{2}$ & $0$ & $+\frac{1}{2}$ & $+\frac{1}{2}$ & $0$ &  $-2$ & $0$  \\
$\mathcal{Q}^1{}_-$ & $- \frac{1}{2}$ & $0$ & $+\frac{1}{2}$ & $+\frac{1}{2}$ & $0$ &  $0$ & $0$ \\
$\mathcal{Q}^2{}_+$ & $+ \frac{1}{2}$ & $0$ & $-\frac{1}{2}$ & $+\frac{1}{2}$& $0$ & $0$ & $+2$ \\
$\mathcal{Q}^2{}_-$ & $- \frac{1}{2}$ & $0$ & $-\frac{1}{2}$ & $+\frac{1}{2}$& $0$ & $+2$ & $+2$ \\
$\tilde{\mathcal{Q}}_{1\dot{+}}$ & $0$ & $+\frac{1 }{2}$ & $-\frac{1}{2}$ & $-\frac{1}{2}$ & $0$ & $+2$ & $0$\\
$\tilde{\mathcal{Q}}_{1\dot{-}}$ & $0$ & $-\frac{1 }{2}$ & $-\frac{1}{2}$ & $-\frac{1}{2}$ & $0$  & $+2$ & $+2$\\
$\tilde{\mathcal{Q}}_{2\dot{+}}$ & $0$ & $+ \frac{1}{2}$ & $+\frac{1}{2}$ &$-\frac{1}{2}$& $0$ & $0$ & $-2$ \\
$\tilde{\mathcal{Q}}_{2\dot{-}}$ & $0$ & $- \frac{1}{2}$ & $+\frac{1}{2}$ &$-\frac{1}{2}$& $0$ & $0$ & $0$\\
\hline
$\mathcal{Q}^3{}_+$ &$+ \frac{1}{2}$& $0$ & $0$ & $-\frac{1}{2}$ & $+1$ & $0$ & $0$\\
$\mathcal{Q}^3{}_-$ &$- \frac{1}{2}$& $0$ & $0$ & $-\frac{1}{2}$ & $+1$ & $+2$ & $0$ \\
$\tilde{\mathcal{Q}}_{3\dot{+}}$ & $0$ & $+ \frac{1}{2}$ & $0$ & $+\frac{1}{2}$ & $-1$ & $0$ & $0$\\
$\tilde{\mathcal{Q}}_{3\dot{-}}$ & $0$ & $- \frac{1}{2}$ & $0$ & $+\frac{1}{2}$ & $-1$ & $0$ & $+2$\\
\hline
\end{tabular}
\caption{The quantum numbers of supercharges}
\label{table:supercharges}
\end{center}
\end{table}

The 4d \Nequals3 superconformal algebra is $\mathfrak{su}(2,2|3)$, whose generators and (anti-)commutation relations in our notation are summarized in appendix \ref{app:SCA}. 
In particular, the fermionic generators are $\mathcal{Q}^I{}_\alpha,\,\tilde{\mathcal{Q}}_{I\dot{\alpha}},\,\mathcal{S}_I{}^{\alpha},\,\tilde{\mathcal{S}}^{I\dot{\alpha}}$ for $\alpha=\pm,\;\dot{\alpha} = \dot{\pm}$ and $I=1,2,3$. 
This algebra has an \Nequals2 superconformal subalgebra containing $\mathcal{Q}^i_\alpha,\,\tilde{\mathcal{Q}}_{i\dot{\alpha}},\,\mathcal{S}_i{}^{\alpha},\,\tilde{\mathcal{S}}^{i\dot{\alpha}}$ for $i=1,2$, whose R-symmetry is $\mathfrak{u}(2)$ generated by $\mathcal{R}^i{}_j$ for $i,j=1,2$. The $\mathfrak{su}(2)_R$ and $\mathfrak{u}(1)_r$ charges are respectively given by
\begin{align}
R\equiv \frac{1}{2}\left(\cR^1{}_1 - \cR^2{}_2\right),\qquad r\equiv \cR^1{}_1 + \cR^2{}_2~.
\label{eq:N=2_R-charges}
\end{align}
The $\cR^i{}_j$ for $i,j=1,2$ and $\cR^3{}_3$ generate an $\mathfrak{su}(2)_R\oplus \mathfrak{u}(1)_r \oplus \mathfrak{u}(1)_F$ subalgebra of $\mathfrak{u}(3)$. Here we take $\mathfrak{u}(1)_F$ to be generated by
\begin{align}
F \equiv 2 \mathcal{R}^3{}_3 + r~,
\label{eq:flavor}
\end{align}
so that our $\mathcal{N}{=}2$ supercharges are neutral under $\mathfrak{u}(1)_F$. 
From the $\mathcal{N}{=}2$ viewpoint, $F$ is a flavor charge.

The quantum numbers of the supercharges are listed in Table \ref{table:supercharges} together with the eigenvalues of the following linear combinations of charges:
\begin{align}
\delta_1 &\equiv \frac{1}{2}\{\cQ^1{}_-,\,(\cQ^1{}_-)^\dagger\} = E-2j_1-2R-r~,
\nonumber\\
 \delta_2 &\equiv \frac{1}{2}\{\tcQ_{2\dot{-}},\, (\tcQ_{2\dot{-}})^\dagger\}=E-2j_2-2R+r~,
\label{eq:delta12}
\end{align}
where $E$ is the scaling dimension and $j_1,j_2$ are the $\mathfrak{so}(4)$ spins such that $\cM_+{}^+ = j_1,\,\cM^{\dot{+}}{}_{\dot{+}} = -j_2$.
We will use the above two linear combinations of charges to discuss, in the next sub-section, the 2d chiral algebras associated in the sense of \cite{Beem:2013sza} to \Nequals3 SCFTs.

The anti-commutation relations \eqref{eq:QS} imply various unitarity bounds on operators. In particular, the presence of the third set of supercharges implies the following unitarity bounds
\begin{align}
\frac{1}{2}\{\cQ^3{}_\pm,\,(\cQ^3{}_\pm)^\dagger\} &= E\pm 2j_1 - F+r\geq 0~,
\nonumber\\
\frac{1}{2}\{\tcQ_{3\dot{\pm}},\,(\tcQ_{3\dot{\pm}})^\dagger\} &= E\pm 2j_2 +F-r\geq 0~.
\label{eq:N=3_unitarity}
\end{align}
This particularly means that any scalar operator should have $E\geq |F-r|$.

\subsubsection{Higgs branch operators}

The \Nequals3 unitarity bounds \eqref{eq:N=3_unitarity} are further simplified for the special set of operators called Higgs branch operators. They are defined as local operators annihilated by all of $\cQ^1{}_\alpha,\, (\cQ^1{}_\alpha)^\dagger$ and $\tilde{\cQ}_{2\dot{\alpha}},\,(\tilde{\cQ}_{2\dot{\alpha}})^\dagger$ for $\alpha=\pm $ and $\dot\alpha=\dot\pm$. Since they saturate the following bounds
\begin{align}
\frac{1}{2}\{\cQ^1{}_\pm,\, (\cQ^1{}_\pm)^\dagger\} &= E \pm 2j_1 -2R-r \geq 0~,
\nonumber\\
\frac{1}{2}\{\tcQ{}_{2\dot{\pm}},\, (\tcQ{}_{2\dot{\pm}})^\dagger\} &= E \pm 2j_2 - 2R+r\geq 0~,
\label{eq:Higgs_unitarity}
\end{align}
they are conformal primaries with $E=2R$ and $j_1=j_2=r=0$. For these operators, the \Nequals3 unitarity bounds \eqref{eq:N=3_unitarity} reduce to 
\begin{align}
E\geq |F|~.
\label{eq:N=3_unitary_Higgs}
\end{align}
Moreover, Higgs branch operators are annihilated by all of $\cS_{I}{}^\alpha = (\cQ^{I}{}_\alpha)^\dagger$ and $\tcS^{I\dot{\alpha}} = (\tcQ_{I\dot{\alpha}})^\dagger$ for $I=1,2,3$. Indeed $\cS_1{}^\alpha$ and $\tcS^{2\dot{\alpha}}$ annihilate them by definition while the action of the other $\cS_{I}{}^\alpha,\,\tcS^{I\dot{\alpha}}$ on Higgs branch operators breaks one of the unitarity bounds in \eqref{eq:Higgs_unitarity}. Therefore any Higgs branch operator is an \Nequals3 superconformal primary.

For rank-1 \Nequals3 SCFTs, we have seen in Sec.~\ref{moduli} that there are three generators of the Higgs branch chiral ring, $W^+,\,W^-$ and $J$. Since they respectively have $(E,F) = (\ell,\ell),(\ell,-\ell)$ and $(2,0)$, the $W^\pm$ saturate the \Nequals3 unitarity bound \eqref{eq:N=3_unitary_Higgs} but $J$ does not. In particular, $W^+$ is annihilated by $\cQ^1{}_\alpha,\,\tcQ_{2\dot\alpha},\,\cQ^3{}_\alpha$ (and their conjugates), while $W^-$ is annihilated by $\cQ^1{}_\alpha,\,\tcQ{}_{2\dot\alpha},\,\tcQ_{3\dot\alpha}$ (and their conjugates).

\subsubsection{Coulomb branch operators}

Let us next consider Coulomb branch operators, which are defined as {\it scalar} local operators annihilated by all of $\tcQ{}_{1\dot\alpha},\,(\tcQ{}_{1\dot\alpha})^\dagger$ and $\tcQ{}_{2\dot\alpha},\,(\tcQ{}_{2\dot\alpha})^\dagger$ for $\dot\alpha=\dot\pm$. They saturate the following unitarity bounds
\begin{align}
\frac{1}{2}\{\tcQ{}_{1\dot\pm},\, (\tcQ{}_{1\dot\pm})^\dagger\} &= E \pm 2j_2 +2R+r~ \geq 0~,
\nonumber\\
\frac{1}{2}\{\tcQ{}_{2\dot\pm},\, (\tcQ{}_{2\dot\pm})^\dagger\} &= E \pm 2j_2 -2R+r~ \geq 0~,
\label{eq:Coulomb}
\end{align}
and therefore have $E=-r$ and $R=0$ in addition to $j_1 = j_2 =0$.\footnote{Here $j_1=0$ follows from the fact that Coulomb branch operators are, by definition, scalars. The absence of local operators saturating these bounds with $j_1\neq 0$ in a large class of 4d \Nequals2 SCFTs were discussed in \cite{Buican:2014qla}.} Moreover, they are neutral under any \Nequals2 flavor symmetry \cite{Buican:2013ica, Buican:2014qla}, which implies they have $F=0$. Then we see that Coulomb branch operators saturate the first unitarity bound in \eqref{eq:N=3_unitarity}, and therefore are annihilated not only by $\tcQ_{1\dot\alpha},\,\tcQ_{2\dot\alpha}$ (and their conjugates) but also by $\cQ^3{}_{\alpha}$ (and its conjugate).\footnote{The conjugates of Coulomb branch operators have $E=r$ and saturate the second bound in \eqref{eq:N=3_unitarity}.} From the unitarity bounds \eqref{eq:N=3_unitarity} and \eqref{eq:Coulomb}, we also see that they are \Nequals3 superconformal primaries.

For rank-1 \Nequals3 SCFTs, there is only one Coulomb branch operator $u$. Its $E=-r$ is determined by the fact that $u$ can be regarded as a Higgs branch operator with respect to another set of \Nequals2 supercharges, say $\cQ^3{}_\alpha,\,\tcQ_{3\dot\alpha},\,\cQ^2{}_\alpha,\,\tcQ_{2\dot\alpha}$. With this new choice of \Nequals2 symmetry, $\cQ^1{}_\alpha$ and $\tcQ_{1\dot\alpha}$ are regarded as the ``third'' set of supercharges. Since $u$ is annihilated by $\cQ^3{}_\alpha,\,\tcQ_{2\dot\alpha}$ and their conjugates, it is indeed regarded as a Higgs branch operator with respect to the new \Nequals2 supersymmetry. Moreover, $u$ is annihilated by the anti-chiral part of the ``third'' set of supercharges, $\tcQ{}_{1\dot\alpha}$. This implies that $u$ is mapped to $W^-$ (and vice versa) by exchanging $(\cQ^1{}_\alpha,\,\tcQ_{1\dot\alpha})$ and $(\cQ^3{}_\alpha,\,\tcQ_{3\dot\alpha})$. Since this exchanging is a part of the $\U(3)_R$ symmetry of the theory, we see that the scaling dimension of $u$ is given by $\Delta(u) = \Delta(W^-) = \ell$.\footnote{Exchanging $(\cQ^2{}_\alpha,\,\tcQ_{2\dot\alpha})$ and $(\cQ^3{}_\alpha,\,\tcQ_{3\dot\alpha})$ maps $u$ to the conjugate of $W^+$ and vice versa.}

More generally, for any 4d \Nequals3 SCFT, exchanging  $(\cQ^1{}_\alpha,\,\tcQ_{1\dot\alpha})$ and $(\cQ^3{}_\alpha,\,\tcQ_{3\dot\alpha})$ maps a Coulomb branch operator to a Higgs branch operator saturating the second unitarity bound in \eqref{eq:N=3_unitarity}. Since $E=2R$ for Higgs branch operators is an integer, we see that $E=-r$ for Coulomb branch operators is always an integer for any 4d \Nequals3 SCFT.\footnote{This also follows from the fact that $\cR^2{}_2 - \cR^3{}_3 = r- R-\frac{F}{2}$ has only integer eigenvalues as $\cR^1{}_1 - \cR^2{}_2 = 2R$.}

\subsection{Identifying the 2d $\mathcal{N}{=}2$ super Virasoro multiplet}
\label{subsec:superVirasoro}

In this sub-section, we show that the 2d chiral algebra $\chi[\cT]$ corresponding in the sense of \cite{Beem:2013sza} to any 4d \Nequals3 SCFT, $\cT$, contains an \Nequals2 super Virasoro algebra.\footnote{This was also noticed by O. Aharony, M. Evtikhiev and R. Yacoby (unpublished).}

First of all, let us recall that Schur operators are defined as local operators with $\delta_1=\delta_2=0$, where $\delta_1,\,\delta_2$ are defined in \eqref{eq:delta12}. Their quantum numbers satisfy
\begin{align}
 j_1 + j_2 = E-2R~,\qquad j_1-j_2 = -r~.
\label{eq:Schur}
\end{align}
The unitarity implies that they are operators annihilated by $\cQ^1{}_-,\,(\cQ^1{}_-)^\dagger$ and $\tcQ_{2\dot{-}},\,(\tcQ_{2\dot{-}})^\dagger$. Any local operator which is not a Schur operator has $\delta_1>0$ or $\delta_2>0$. It was shown in \cite{Beem:2013sza} that the space of Schur operators in any 4d \Nequals2 SCFT has a structure of a 2d chiral algebra. In particular, every 4d Schur operator $\cO$ maps to a 2d local operator $\chi[\cO]$ with 2d chiral operator product expansions (OPEs) determined by 4d OPEs. The 2d chiral algebra always contains a Virasoro subalgebra with the identification 
\begin{align}
L_0 = E-R~.
\label{eq:hol-dim}
\end{align}
The  general discussion for \Nequals2 SCFTs in \cite{Beem:2013sza} tells us that our theory $\cT$ has at least the following bosonic Schur operators: 
\begin{itemize}
\item The highest weight component of the $\SU(2)_R$ current, $\cJ^{11}_{+\dot{+}}$, with $E=3,\,R=1$ and $F=0$.\footnote{Here, we follow the convention of \cite{Beem:2013sza}. Namely, $\cJ_{+\dot{+}}^{11}$ is the highest weight $\su(2)_R\oplus \so(4)$ component of the $\SU(2)_R$ current $\cJ_{\alpha\dot{\alpha}}^{ij}$.} 
The corresponding 2d operator 
\begin{align}
T\equiv  \chi[\cJ^{11}_{+\dot{+}}]
\end{align}
is the 2d stress tensor.

\item The highest weight component of the $\U(1)_F$ moment map operator, $J^{11}$, with $E=2,\,R=1$ and $F=0$. This is a Higgs branch operator in the sense of Sec.~\ref{subsubsec:unitarity}, and was denoted by $J$ in Sec.~\ref{moduli}.
The corresponding 2d operator 
\begin{align}
J \equiv \chi[J^{11}]
\end{align}
is an affine $\U(1)$ current.
\end{itemize}
Other bosonic Schur operators will be discussed in the next sub-section.

Since our theory $\cT$ has $\mathcal{N}{=}3$ symmetry, there are extra supercharges $\cQ^3{}_\alpha,\,\tcQ_{3\dot{\alpha}}$. From Table~\ref{table:supercharges}, we see that $\cQ^3{}_+$ and $\tcQ_{3\dot{+}}$ have $\delta_1=\delta_2=0$ and therefore act on the space of Schur operators.\footnote{On the other hand, $\cQ^3{}_-$ and $\tcQ_{3\dot{-}}$ have either $\delta_1>0$ or $\delta_2>0$, and therefore their actions cannot create any Schur operator. They map any local operator to a non-Schur operator or zero.}
This means that fermionic Schur operators are created by acting $\cQ^3{}_+$ and $\tcQ_{3\dot{+}}$ on the above bosonic ones.
For example, $\cQ^3{}_+ J^{11}$ and $\tcQ_{3\dot{+}}J^{11}$ are two fermionic Schur operators,
which are non-vanishing due to the unitarity bounds \eqref{eq:N=3_unitarity}.\footnote{In the language of \cite{Dolan:2002zh}, these operators are respectively in the $\mathcal{D}_{\frac{1}{2}(0,0)}$ and the $\bar{\mathcal{D}}_{\frac{1}{2}(0,0)}$ multiplets. } Moreover, they are conformal primaries because $J^{11}$ is an \Nequals3 superconformal primary as shown in Sec.~\ref{subsubsec:unitarity}. Then, as shown in appendix \ref{app:computation2}, the corresponding 2d operators
\begin{align}
G \equiv  \chi[\cQ^3{}_+J^{11}]~,\qquad \bar{G} \equiv \chi[\tcQ_{3\dot{+}}J^{11}]
\label{eq:G}
\end{align}
are Virasoro primaries. From \eqref{eq:hol-dim}, we see that their holomorphic dimension is $\frac{3}{2}$. Moreover, $G$ and $\bar{G}$ respectively have charge $+1$ and $-1$ under $J$ since $\cQ^3{}_+$ and $\tcQ_{3\dot+}$ have $\U(1)_F$ charge $\pm1$.

Let us next consider $\{\cQ^3{}_+,\,\tcQ{}_{3\dot{+}}\}J^{11}$ and $[\cQ^3{}_+,\,\tcQ_{3\dot{+}}]J^{11}$. While the former is a conformal descendant of $J^{11}$, the latter is a bosonic Schur operator with $E=3,\,R=1$ and moreover is a conformal primary. According to \cite{Dolan:2002zh, Gadde:2011uv, Beem:2013sza}, the only such Schur operator is the highest weight component of the $\SU(2)_R$ current, $\cJ^{11}_{+\dot{+}}$, in the stress tensor multiplet. Assuming the unique stress tensor in $\cT$, we conclude that
\begin{align}
\cJ_{+\dot{+}}^{11} = \frac{1}{2}[\cQ^3{}_+,\,\tcQ_{3\dot{+}}]J^{11}.
\end{align}
More generally we identify $\cJ_{\alpha\dot{\alpha}}^{ij} = \frac{1}{2}[\cQ^3{}_\alpha,\,\tcQ_{3\dot{\alpha}}]J^{ij}$.

Hence, the four Schur operators $J^{11},\,\cQ^3{}_+J^{11},\,\tcQ_{3\dot{+}}J^{11}$ and $\cJ^{11}_{+\dot{+}}$ are in the same \Nequals3 superconformal multiplet as the stress tensor.\footnote{Further actions of $\cQ^3{}_+$ or $\tcQ{}_{3\dot{+}}$ on these operators do not create any new Schur operators up to their conformal descendants.} This means that the corresponding 2d chiral operators $J,\,G,\,\bar{G}$ and $T$ are also in a 2d super multiplet. 
It is a standard fact that in 2d, the energy momentum tensor $T$, a $\U(1)$ current $J$, and two fermionic dimension 3/2 currents $G,\,\bar{G}$ of $\U(1)$ charge $\pm1$ necessarily form the \Nequals2 super Virasoro algebra.
Therefore, we see that the 2d chiral algebra $\chi[\mathcal{T}]$ associated in the sense of \cite{Beem:2013sza} to a 4d \Nequals3 SCFT contains the \Nequals2 super Virasoro algebra.
The 2d central charge is \begin{equation}
c_{2d}=-12c_{4d} 
\end{equation} as in \cite{Beem:2013sza}. 

The $\mathfrak{sl}(2|1)$ subalgebra of the \Nequals2 super Virasoro algebra can be explicitly seen in the \Nequals3 superconformal algebra. Indeed, $L_0, \,L_{\pm1}$ was identified as $L_{-1} = \frac{1}{2}\cP_{+\dot+},\,L_1= \frac{1}{2}\cK^{\dot++}$ and $L_0=E-R$ in \cite{Beem:2013sza},\footnote{The extra factor of $\frac{1}{2}$ comes from our different normalization of $\cP_{\alpha\dot\alpha}$ and $\cK^{\dot\alpha\alpha}$.} and our identification \eqref{eq:G} means
\begin{align}
G_{-\frac{1}{2}} = \frac{1}{2}\cQ^3{}_+,\quad \bar{G}_{-\frac{1}{2}} = \frac{1}{2}\tcQ_{3\dot+},\quad G_{\frac{1}{2}} = \frac{1}{2}\tcS{}^{3\dot+},\quad \bar{G}_{\frac{1}{2}}=\frac{1}{2}\cS_3{}^+,\quad J_0=F~.
\end{align}
It is then straightforward to show that, under these identifications, $L_0,\,L_{\pm 1},\,J_0$ and $G_{\pm \frac{1}{2}},\,\bar{G}_{\pm \frac{1}{2}}$ generate a subalgebra of $\mathfrak{su}(2,2|3)$ which acts as $\mathfrak{sl}(2|1)$ on the space of Schur operators.

\subsection{2d operators corresponding to Higgs branch operators}
\label{subsec:higgsop}

In addition to the above Schur operators, the Higgs branch operators are all Schur operators.
We here show the following two statements:
\begin{enumerate}
\item For any Higgs branch operator $\mathcal{O}$, $\chi[\mathcal{O}]$ is a superprimary operator.
\item For any Higgs branch operator $\mathcal{O}$ with $E= \pm F$, $\chi[\mathcal{O}]$ is a(n) (anti-)chiral superprimary.
\end{enumerate}
In the next section, we will  use the second statement to identify the 2d chiral algebras corresponding in the sense of \cite{Beem:2013sza} to rank-1 \Nequals3 SCFTs.

Let us first show the first statement. Suppose that $\mathcal{O}$ is a Higgs branch operator. Since $\mathcal{O}$ is a Hall-Littlewood operator in the language of \cite{Gadde:2011uv, Beem:2013sza}, $\chi[\mathcal{O}]$ is a Virasoro primary in two dimensions (as shown in Sec.~3.2.4 of \cite{Beem:2013sza} and reviewed in appendix \ref{app:computation2}). Therefore we only need to show that $\chi[\mathcal{O}]$ is annihilated by $G_{n+\frac{1}{2}}$ for $n\geq 0$. Since $\mathcal{O}$ is an $\mathcal{N}{=}3$ superconformal primary as shown in Sec.~\ref{subsubsec:unitarity}, it is annihilated by $(\mathcal{Q}^{3}{}_+)^\dagger,\, (\tilde{\mathcal{Q}}_{3\dot{+}})^\dagger$. This means that $\chi[\mathcal{O}]$ is annihilated by $G_{\frac{1}{2}}$ and $\bar{G}_{\frac{1}{2}}$. Therefore, for all $n\geq 2$,
\begin{align}
G_{n+\frac{1}{2}} = \frac{2}{n-1}[L_n,\, G_{\frac{1}{2}}]~,\qquad \bar{G}_{n+\frac{1}{2}} = \frac{2}{n-1}[L_n,\, \bar{G}_{\frac{1}{2}}]~
\end{align}
also annihilate $\chi[\mathcal{O}]$. Finally, 
\begin{align}
G_{\frac{3}{2}}\,\chi[\mathcal{O}] = \frac{2}{3}L_2G_{-\frac{1}{2}}\,\chi[\mathcal{O}]~,
\qquad 
\bar{G}_{\frac{3}{2}}\,\chi[\mathcal{O}] = \frac{2}{3}L_2\bar{G}_{-\frac{1}{2}}\,\chi[\mathcal{O}]~,
\label{eq:GO}
\end{align}
are vanishing because  $G_{-\frac{1}{2}}\chi[\mathcal{O}]$ and $\bar{G}_{-\frac{1}{2}}\chi[\mathcal{O}]$ are Virasoro primaries (see appendix \ref{app:computation2}). Hence, $\chi[\cO]$ is a superprimary in two dimensions.

Let us next consider the second statement. Note that the requirement $E=\pm F$ is precisely the condition that one of the unitarity bounds in \eqref{eq:N=3_unitarity} is saturated since $j_{1,2}=r=0$ here. 
Therefore, if a Higgs branch operator, $\mathcal{O}$, has $E=+ F$ (or $E=-F$), then $\mathcal{O}$ is annihilated by $\mathcal{Q}^3{}_\alpha$ (or $\tilde{\mathcal{Q}}_{3\dot{\alpha}}$). This particularly means that the corresponding 2d operator, $\chi[\mathcal{O}]$, is annihilated by $G_{-\frac{1}{2}}$ (or $\bar{G}_{-\frac{1}{2}}$). Thus, we see that any Higgs branch operator with $E=F$ (or $E=-F$) maps to an anti-chiral (or chiral) superprimary in two dimensions.

\section{Construction of the associated 2d chiral algebras}\label{jacobi}
Based on the properties we uncovered in the previous section,
here we proceed to the construction of the 2d chiral algebras associated in the sense of \cite{Beem:2013sza} to the 4d \Nequals3 rank-1 superconformal field theories, whose moduli space is of the form $\bC^3/\bZ_\ell $ where $\ell=1,2,3,4,6$. 

As shown in Sec.~\ref{subsec:superVirasoro}, the 2d chiral algebra has an \Nequals2 super Virasoro algebra as a subalgebra. 
In addition, the Higgs branch operators as 4d \Nequals2 theory give rise to generators of the 2d chiral algebra, as was shown in \cite{Beem:2013sza}. 
In our setup, the Higgs branch operators in 4d are generated by $W_+$, $W_-$ and $J$, whose dimensions are $\ell$, $\ell$, $2$ and the $\U(1)$ charges are $\ell$, $-\ell$, $0$ respectively, with one relation \begin{equation}
W_+ W_- \propto J^\ell .
\end{equation}
As shown in Sec.~\ref{subsec:superVirasoro}, $\chi[J]$ is the bottom component of the super energy momentum tensor, and $\chi[W_+]$ ($\chi[W_-]$) is a(n) (anti)chiral primary of dimension $\ell/2$.
Below, we use the following shorthand notations for them: \begin{equation}
J:=\chi[J],\qquad W:=\chi[W_+],\qquad \bar W:=\chi[W_-].
\end{equation} 

In the cases studied previously in the literature e.g.~\cite{Beem:2013sza,Beem:2014rza,Lemos:2014lua}, it was often the case that the entire 2d chiral algebras were generated by taking repeated operator product expansions of the Higgs branch operators. 
We use this empirical feature as a working hypothesis and will find out that it leads to a consistent answer. 
As it is important, let us record here our \textsc{Assumption:}
\begin{quotation}
\noindent The 2d chiral algebra is generated by the \Nequals2 super Virasoro multiplet $J$, a bosonic chiral primary $W$ and a bosonic antichiral primary $\bar W$, both of dimension $\ell/2$.
\end{quotation}

We will see below that for $\ell=3$, this assumption uniquely fixes $c_\text{2d}$ to be $-15$,
consistent with the 4d central charge $c_\text{4d}=(2\ell-1)/4$ derived in \eqref{centralcharges}
with the standard mapping $c_\text{2d}=-12 c_\text{4d}$.
Furthermore, we see that the construction automatically leads to a null relation of the form \begin{equation}
W \bar W \propto J^3 + (\text{composite operators constructed from $J$ and (super)derivatives}),
\end{equation}  reproducing the Higgs branch relation.

Similarly, for $\ell=4$, the allowed $c_\text{2d}$ are $-21$, $-9$ and $12$, with the Higgs branch relation reproduced  for $c_\text{2d}=-21$, and for $\ell=6$, the allowed $c_\text{2d}$ are $-33$, $-15$ and $18$, with the Higgs branch relation reproduced  for $c_\text{2d}=-33$.

Before proceeding, we note that the 2d chiral algebras satisfying the assumption above were constructed  in \cite{Odake:1988bh,Inami:1989yi} for $\ell=3$ but with $W$ and $\bar W$ implicitly taken to be fermionic.
This choice was more natural for a 2d unitary algebra, since the spin of $W$ and $\bar W$ is half-integral. 
In this case the allowed central charge was $c_\text{2d}=+9$. 
The 2d chiral algebras for $\ell=4$ and $\ell=6$ with bosonic $W_\pm$ were constructed in \cite{Blumenhagen:1992sa}, with the allowed central charges as listed above.
The null relation leading to the Higgs branch relation was not studied there. 

\subsection{Conventions}
In the computations below, we use the 2d \Nequals2 holomorphic superspace, where the coordinate $Z$ consists of the bosonic coordinate  $z$ and the fermionic coordinates $\theta$ and $\bar\theta$.
We mostly follow the convention of Krivonos and Thielemans \cite{Krivonos:1995bk}, where the Mathematica package \texttt{SOPEN2defs} we will use to compute the \Nequals2 superconformal operator product expansion was developed and described.\footnote{ Note that they called the operators satisfying $\bar \cD W$ antichiral primary, but we call such operators chiral primary and vice versa. They also  had a typo in their super OPE of the superconformal algebra in their (7), where $c/4$ should be $c/3$.}

We define the superderivatives to be \begin{equation}
\cD = \partial_{\theta} -\frac12 \bar\theta \partial_z,\qquad
\bar\cD = \partial_{\bar\theta} -\frac12 \theta \partial_z,\qquad
\end{equation} Then a chiral superfield $W$ and an antichiral $\bar W$ satisfy \begin{equation}
\bar\cD W=0, \qquad \cD \bar W=0
\end{equation} respectively. 
The operator product expansions can be usefully done using covariant combinations \begin{equation}
Z_{12}=z_1-z_2+\frac12(\theta_{1}\bar\theta_{2}-\theta_{2}\bar\theta_{1}),\qquad
\theta_{12}=\theta_{1}-\theta_{2},\qquad
\bar\theta_{12}=\bar\theta_{1}-\bar\theta_{2}.
\end{equation}

Then the energy momentum superfield $J(Z)$ has the operator product expansion \begin{equation}
J(Z_1)J(Z_2)\sim \frac{c/3+\theta_{12}\bar\theta_{12} J}{Z_{12}^2}
+\frac{-\theta_{12}\cD J+\bar\theta_{12}\bar\cD  J +\theta_{12}\bar\theta_{12} \partial J }{Z_{12}}
\label{JJOPE}
\end{equation} and a superprimary $\mathcal{O}$ with dimension $\Delta$ and $\U(1)$ charge $F$ has the operator product expansion with $J$ given by \begin{equation}
J(Z_1)\mathcal{O}(Z_2) \sim \Delta\frac{\theta_{12}\bar\theta_{12} \mathcal{O}}{Z_{12}^2}
+\frac{-F \mathcal{O} -\theta_{12}\cD \mathcal{O}+\bar\theta_{12}\bar\cD  \mathcal{O} +\theta_{12}\bar\theta_{12} \partial \mathcal{O} }{Z_{12}}.
\label{superprimaryOPE}
\end{equation} 
Here, in the equations \eqref{JJOPE}and \eqref{superprimaryOPE} and below, the operators on the right hand sides of the operator product expansions are always taken to be at $Z=Z_2$.

In our convention the (anti)chiral primaries are those with $\Delta=F/2$ ($\Delta=-F/2$).
Note that our 2d algebra is not unitary, and therefore, $\Delta=F/2$ does not immediately imply that the antichiral derivative to vanish.
Rather,  we use the fact that $W$ and $\bar W$ come from 4d Higgs branch operators $W_+$ and $W_-$ of an \Nequals3 theory to conclude that they are (anti)chiral primaries.

The normal ordered product of two operators $\mathcal{O}_1$ and $\mathcal{O}_2$ is defined as the constant term, i.e.~the term without any power of  $\theta_{12}$, $\bar\theta_{12}$ or $Z_{12}$ in the operator product expansion of $\mathcal{O}_1$ and $\mathcal{O}_2$. 
Note that this does not always agree with the normal ordered product of two operators defined as the constant part of the operator product expansion of the bottom components on the non-superspace parametrized only by $z$.  
The normal ordered product of more than two operators are defined by
 recursively taking the operator product expansions from the right, i.e.~ $\mathcal{O}_1\mathcal{O}_2 \mathcal{O}_3 \cdots = (\mathcal{O}_1(\mathcal{O}_2(\mathcal{O}_3\cdots )))$.

\subsection{Strategy}
Our computational strategy is quite simple. We first require the operator product expansions of $J$ with itself \eqref{JJOPE},
and that $W$, $\bar W$ have the operator product expansions with respect to $J$ given by \eqref{superprimaryOPE} where $\Delta=\ell/2$ and $F=\pm \ell$,
and that \begin{equation}
W (Z_1)W (Z_2) \sim \text{regular},\qquad
\bar W (Z_1)\bar W (Z_2) \sim \text{regular}.
\end{equation}
The only operator product expansion that needs to be worked out is that of $W $ and $\bar W $.
 
Our assumption implies that only $J$ and composite operators constructed out of it appear in the singular part of this operator product expansion. 
Demanding that $W (Z_1)\bar W (Z_2)$ to be annihilated by $\bar\cD _1$ and $\cD _2$, we find that it has the form \begin{equation}
W (Z_1)\bar W (Z_2) \sim \sum_{d=1}^\ell  \frac{1}{Z_{12}{}^d}\left(\frac d2\frac{ \theta_{12}\bar\theta_{12}} {Z_{12}}+1+\theta_{12} \cD \right){\mathcal{O}_{\ell-d}}
\label{WWOPE}
\end{equation}
where $\mathcal{O}_d$ is an operator of dimension $d$ constructed out of $J$ and its (super)derivatives.
In particular, $\mathcal{O}_0$ is a constant and $\mathcal{O}_1 \propto J$.
We arbitrarily choose $\mathcal{O}_1=J$ to fix the normalization of $W$ and $\bar W$. 
Demanding the closure of the Jacobi identity among $J$, $W $ and $\bar W $ then fixes all other $\mathcal{O}_d$.
Note that this is just the standard fact that when we fix the normalization of a primary (this time, the identity operator) in an operator product expansion, the contribution of all the descendants are automatically fixed. 
The explicit expressions for $\mathcal{O}_d$ are given in \cite{Blumenhagen:1992sa}.

The only nontrivial procedure is to check the closure of the Jacobi identity among $W $, $W $ and $\bar W $; the analysis of the Jacobi identity for the triple $\bar W $, $\bar W $ and $W $ is similar, thanks to the discrete symmetry exchanging $W$ and $\bar W$.

The computations can be performed easily and quickly using \texttt{SOPEN2defs}, the Mathematica package written by Krivonos and Thielemans \cite{Krivonos:1995bk}.  On  a 2012 notebook computer, the computation time was dominated by the time needed to type expressions into a notebook. The entire computation of Jacobi identities etc.~took at most a few minutes.
The Mathematica notebook detailing the computations below is available as ancillary files on the arXiv page for this paper.

\subsection{Results}
\subsubsection{$\ell=1$}
When $\ell=1$, the operator product expansion \eqref{WWOPE} just means that $W$ and $\bar W$ are free, consisting of two bosons $q$, $\bar q$ of dimension $1/2$ with $q \bar q\sim 1/z$ and two neutral fermions $\lambda$, $\bar\lambda$ of dimension $1$ with $\lambda \bar \lambda \sim 1/z^2$. 
These are as they should be, since the 4d theory itself is free.
We can define $J=W \bar W $ to reproduce the (rather trivial) Higgs branch relation.
This $J$ automatically has the correct operator product expansion \eqref{JJOPE} with $c_\text{2d}=-3$, which agrees with the expected formula $c_\text{2d}=-3(2\ell-1)$.
In fact this case was already essentially discussed in \cite{Beem:2013sza}.

\subsubsection{$\ell=2$}
When $\ell=2$, the operator product expansion \eqref{WWOPE} together with the other operator product expansions mean that $W $, $J$, $\bar W $ generate a small \Nequals4 super Virasoro algebra. 
As such, the operator product expansions close for arbitrary value of $c_\text{2d}$.
Explicitly, we need to choose $\mathcal{O}_0=-c/3$ and $\mathcal{O}_1=J$.

It is still instructive to see when there can be null relations representing the Higgs branch relation $W \bar W  \propto J^2$. 
In the language of the 2d chiral algebra, this should correspond to a null relation of the form \begin{equation}
W \bar W  - (a_1 J^2 + a_2 J'+ a_3 [\cD ,\bar\cD ] J)=0 .\label{foo}
\end{equation}
Demanding that the left hand side  to be an \Nequals2 primary, we find that only two choices are possible: 
\begin{equation}
(c_\text{2d},a_1,a_2,a_3)=(-9,1/2,1,1/2)\quad \text{or} \quad (6,-1,1,0).
\end{equation}
It turns out, however, that only the first choice makes the left hand side of \eqref{foo} to be an \Nequals4 primary. 
For example, with the second choice,  repeated operator product expansions of \eqref{foo} with $W$ leads to an additional null relation $W  ^2 = 0$, which we do not like. 
We see that the Higgs branch relation is only compatible when $c_\text{2d}=-9=3(2\ell-1)$.
Before proceeding, we note that the null relation above for $c_\text{2d}=-9$ leads to new  null operators given by  \begin{equation}
\begin{aligned}
\mathcal{X}&=\cD  \partial W - J\cD W  + 2 (\cD  J) W ,\\
\bar{\mathcal{X}}&=\bar\cD \partial \bar W + J\cD \bar W  - 2 (\bar\cD  J) \bar W .
\end{aligned}\label{null1}
\end{equation}

\subsubsection{$\ell=3$}
The fun starts at $\ell=3$. 
We find that the Jacobi identity for $W $, $W $, $\bar W $ does not close for general values of $c_\text{2d}$, since the failure of the Jacobi identity contains terms proportional to the identity operator.
These terms all vanish when and only when $c_\text{2d}=-15$.
Note that this is exactly what the 4d \Nequals3 analysis dictates: $c_\text{2d}=-12c_\text{4d}=-3(2\ell-1)$.

With this value of the central charge, the $W \bar W $ operator product expansion \eqref{WWOPE} is given by \begin{equation}
\mathcal{O}_0=\frac53,\quad
\mathcal{O}_1=J,\quad
\mathcal{O}_2=\frac14J^2+\frac12\partial J + \frac14 [\cD ,\bar\cD ]J.
\end{equation}
The failure of the Jacobi identity for $W $, $W $, $\bar W $ now  contains only terms proportional to \begin{equation}
\mathcal{X}=\cD  \partial W - J\cD W  + 3 (\cD  J) W \label{null2}
\end{equation} and $\cD \mathcal{X}=-4 \cD  J \cD W$.
One finds that $\mathcal{X}$ happens to be an \Nequals2 superprimary,
so it is possible to impose the null relation $\mathcal{X}=0$ as far as the operator product expansion with $J$ is concerned. 
After imposing this null relation and its \Nequals2 descendants, 
we find that the Jacobi identity for $W $, $W $ and $\bar W $ closes.

Similarly,  we find that the Jacobi identity for $W $, $\bar W $ and $\bar W $ closes after demanding that the composite operator \begin{equation}
\bar{\mathcal{X}}=\bar\cD  \partial \bar W  + J\bar\cD \bar W  - 3 (\bar\cD  J) \bar W \label{null3}
\end{equation} is null.

One further finds that the operator product of $\mathcal{X}$ and $W $ is regular,
while that of $\mathcal{X}$ and $\bar W $ contains operators  whose scaling dimensions are larger than that of $\mathcal{X}$. Similar statements hold $\bar{\mathcal{X}}$. 
This guarantees that $\mathcal{X}$ and $\bar{\mathcal{X}}$ are the operators with lowest dimension among the null states to be removed. 

Another null state is obtained by taking the operator product expansion of $\bar W $ with $\mathcal{X}$, whose coefficient of $\bar\theta_{12}/Z_{12}^2$ is proportional to \begin{equation}
\mathcal{Y}=36 W \bar W 
- (J^3+9 (\partial J) J + 6 J [\cD ,\bar\cD ] J +6 \bar\cD  J \cD J + 6 [\cD ,\bar\cD ] \partial J + 7 \partial^2 J).
\end{equation}
This operator is null, and correctly represents the 4d Higgs branch relation $W  \bar W  \propto J^3$.

\subsubsection{$\ell=4,5,6$}
The analysis for $\ell$ larger than three can similarly be done.
For $\ell=4$, we find that the failure of the Jacobi identity for $W $, $W $, $\bar W $ contains terms proportional to the identity operator times $(c-12)(c+9)(c+21)$.
For each possible case $c=-21$, $-9$ and $12$, 
we find that the Jacobi identity can be satisfied by imposing a null relation.
But we find that the null relation is only consistent with the expected Higgs branch relation  when $c_\text{2d}=-21=-3(2\ell-1)$, again the value that follows from our \Nequals3 analysis.

For $\ell=6$, we find that the values of $c_\text{2d}$ allowed by the closure of the Jacobi identity is $c=-33$, $-15$ and $18$. Again, the null relation is compatible with the Higgs branch relation only for $c_\text{2d}=-33=-3(2\ell-1)$. 
In both cases $\ell=4,6$, we find that the basic null operators are \begin{equation}
\begin{aligned}
\mathcal{X}&=\cD  \partial W - J\cD W  + \ell (\cD  J) W , \\
\bar{\mathcal{X}}&=\bar\cD  \partial \bar W  + J\bar\cD \bar W  - \ell (\bar\cD  J) \bar W .
\end{aligned}\label{nullX}
\end{equation}
Note that the null operators for the cases $\ell=2,3$ are given by the same expressions, see \eqref{null1}, \eqref{null2}, \eqref{null3}. 

We can also analyze the case $\ell=5$. 
Here we find that the Jacobi identities are only consistent for $c_\text{2d}=-27=-3(2\ell-1)$,
and the null relation are generated by the same $\mathcal{X}$ and $\bar{\mathcal{X}}$ given in \eqref{nullX}.
A descendant by $\bar W $ of $\mathcal{X}$ generates a new null relation of the form $W \bar W  \propto J^5 + \text{(operators involving (super)derivatives)}$.
Note that the existence of the $\ell=5$ super W-algebra does \emph{not} contradict with the fact that there should not be the \Nequals3 theory with $\ell=5$ in four dimensions.

The analysis so far suggests that there is a series of super W-algebras generated by the \Nequals2 super Virasoro algebra plus (anti)chiral primaries $W_\pm$ of dimension $\ell/2$ with $c_\text{2d}=-3(2\ell-1)$, with the basic null fields as given in \eqref{nullX}.  
The operator product expansion of $W$ and $\bar W$ has the form \eqref{WWOPE}. A descendant of the null field seems to  automatically  give the relation of the form \begin{equation}
\ell^2 W \bar W = \frac{(2(\ell-1))!}{\ell!} J^\ell  +\text{(descendants)},
\end{equation} where the coefficients are guessed from the examples so far. 
Note that we normalized $W$ and $\bar W$ by demanding $\mathcal{O}_1=J$ in  \eqref{WWOPE}.
It would be interesting, as a question purely in two dimensions, to see whether such a series of 2d chiral algebras indeed exists.

\section*{Acknowledgements}
The authors thank S. Nawata for helping locate the literature on W-algebras,
and T. Nishioka for bringing the paper \cite{Ferrara:1998zt} to the authors' attention.
The authors also thank P. Argyres, M. Lotito, Y. L\"u, M. Martone for instructive discussions on their  papers \cite{Argyres:2015ffa,Argyres:2015gha,ArgyresEtAlToAppear}.
T.~N. would like to thank M. Buican 
and Y.~T. would like to thank O. Aharony
for illuminating discussions in related projects.
The work of T.~N. is partially supported by the Yukawa Memorial Foundation, and 
the work of Y.~T. is partially supported in part by JSPS Grant-in-Aid for Scientific Research No. 25870159,
and  by WPI Initiative, MEXT, Japan at IPMU, the University of Tokyo.

\appendix

\section{The 4d \Nequals3 superconformal algebras}
\label{app:SCA}
 The 4d $\mathcal{N}{=}3$ superconformal algebra is $\mathfrak{su}(2,2|3)$ whose bosonic part is $\mathfrak{su}(2,2)\oplus \mathfrak{u}(3)$. The $\mathfrak{u}(3)$ R-symmetry is generated by $\mathcal{R}_I{}^J$ for $I,J=1,2,3$ subject to
\begin{align}
\left[\mathcal{R}^I{}_J, \mathcal{R}^K{}_L\right] = \delta^K_J \mathcal{R}^I{}_L - \delta^I_L \mathcal{R}^K{}_J~.
\end{align}
The fermionic generators of $\mathfrak{su}(2,2|3)$ are $\mathcal{Q}^I{}_\alpha,\,\tilde{\mathcal{Q}}_{I\dot{\alpha}}$ and $\mathcal{S}_I{}^\alpha,\,\tilde{\mathcal{S}}^{I\dot{\alpha}}$ for $I=1,2,3,\,\alpha= \pm$ and $\dot{\alpha} = \dot{\pm}$. Their R-charges can be read off from
\begin{align}
[\mathcal{R}^I{}_J,\, \mathcal{Q}^K{}_\alpha] = \delta^K_J \mathcal{Q}^I{}_\alpha - \frac{1}{4}\delta^I_J \mathcal{Q}^K{}_\alpha ~ &, \qquad [\mathcal{R}^I{}_J,\,\tilde{\mathcal{Q}}_{K\dot{\alpha}}] = -\delta^I_K\tilde{\mathcal{Q}}_{J\dot{\alpha}}+\frac{1}{4}\delta^I_J\tilde{\mathcal{Q}}_{K\dot{\alpha}}~,
\nonumber\\
[\mathcal{R}^I{}_J,\, \mathcal{S}_K{}^\alpha] = -\delta^I_K \mathcal{S}_J{}^\alpha + \frac{1}{4}\delta^I_J\mathcal{S}_K{}^\alpha~&,\qquad [\mathcal{R}^I{}_J,\, \tilde{\mathcal{S}}^{K\dot{\alpha}}] = \delta^K_J \tilde{\mathcal{S}}^{I\dot{\alpha}} -\frac{1}{4}\delta^I_J\tilde{\mathcal{S}}^{K\dot{\alpha}}~.
\label{eq:R-charges}
\end{align}
The anti-commutation relations among $\mathcal{Q}^I{}_\alpha,\,\tilde{\mathcal{Q}}_{I\dot{\alpha}},\,\mathcal{S}_I{}^\alpha,\,\tilde{\mathcal{S}}^{I\dot{\alpha}}$ are given by
\begin{align}
\{\mathcal{Q}^I{}_\alpha,\, \mathcal{S}_J{}^{\beta}\}= 2\delta^I_J\delta_\alpha^\beta \mathcal{H} + 4\delta^I_J \mathcal{M}_\alpha{}^\beta -4\delta_\alpha^\beta \mathcal{R}^I{}_J~,& \qquad \{\mathcal{Q}^I{}_\alpha,\,\tilde{\mathcal{Q}}_{J\dot{\alpha}}\} = 2\delta^I_J\mathcal{P}_{\alpha\dot{\alpha}}~,
\nonumber\\
\{\tilde{\mathcal{S}}^{I\dot{\alpha}},\, \tilde{\mathcal{Q}}_{J\dot{\beta}}\} = 2\delta^I_J\delta^{\dot{\alpha}}_{\dot{\beta}} \mathcal{H} -4\delta^I_J \tilde{\cM}^{\dot{\alpha}}{}_{\dot{\beta}} + 4\delta^{\dot{\alpha}}_{\dot{\beta}} \mathcal{R}^I{}_J~,&\qquad \{\tilde{\mathcal{S}}^{I\dot{\alpha}},\,\mathcal{S}_J{}^\alpha\} = 2\delta^I_J\mathcal{K}^{\dot{\alpha}\alpha}~.
\label{eq:QS}
\end{align}
Here $\mathcal{H}$ is the Hamiltonian whose eigenvalue is the scaling dimension, and $\mathcal{M}_\alpha{}^{\beta},\,\tilde{\mathcal{M}}^{\dot{\alpha}}{}_{\dot{\beta}}$ are generators of $\mathfrak{so}(4)$ subalgebra of $\mathfrak{su}(2,2)$. They satisfy
\begin{align}
[\cH,\,\cQ^I{}_\alpha] = \frac{1}{2}\cQ^I_\alpha~,\quad [\cH,\,\tcQ_{I\dot\alpha}] =\frac{1}{2}\tcQ_{\dot\alpha}~,&\quad [\cH,\,\cS_I{}^\alpha] = -\frac{1}{2}\cS_I{}^\alpha~,\quad [\cH,\,\tcS^{I\dot\alpha}]=-\frac{1}{2}\tcS^{I\dot\alpha}~,
\nonumber\\
[\cM_\alpha{}^\beta,\,\cQ^{I}{}_\gamma]=\delta^\beta_\gamma \cQ^{I}{}_\alpha - \frac{1}{2}\delta^\beta_\alpha \cQ^I{}_\gamma~,&\quad [\cM_\alpha{}^\beta,\cS_I{}^\gamma] = -\delta^{\gamma}_\alpha S_I{}^\alpha + \frac{1}{2}\delta^\beta_\alpha \cS_I{}^\gamma~,
\nonumber\\
[\tilde{\cM}^{\dot\alpha}{}_{\dot\beta},\,\tcQ_{I\dot\gamma}]= -\delta^{\dot\alpha}_{\dot\gamma} \tcQ_{I\dot\beta} + \frac{1}{2}\delta^{\dot\alpha}_{\dot\beta} \tcQ_{I\dot\gamma}~,&\quad [\tilde{\cM}^{\dot\alpha}{}_{\dot\beta},\,\tcS^{I\dot\gamma}]= \delta^{\dot\gamma}_{\dot\beta} \tcS^{I\dot\alpha} - \frac{1}{2}\delta^{\dot\alpha}_{\dot\beta} \tcS^{I\dot\gamma}~.
\end{align}
On the other hand, $P_{\alpha\dot{\alpha}}$ and $\mathcal{K}^{\dot{\alpha}\alpha}$ have the following commutation relations with the supercharges:
\begin{align}
[\mathcal{K}^{\dot{\alpha}\alpha},\,\mathcal{Q}^{I}{}_{\beta}] = 2\delta^{\alpha}_{\beta}\,\tilde{\mathcal{S}}^{I\dot{\alpha}}~,& \qquad [\mathcal{K}^{\dot{\alpha}\alpha}\tilde{\mathcal{Q}}_{I\dot{\beta}}] = 2\delta^{\dot{\alpha}}_{\dot{\beta}}\, \mathcal{S}_{I}{}^\alpha~,
\nonumber\\
[{\cP}_{\alpha\dot\alpha},\,\cS_I{}^\beta] = -2\delta^{\beta}_{\alpha}\tcQ_{I\dot\alpha}~,&\qquad [\cP_{\alpha\dot\alpha},\,\tcS^{I\dot\beta}] = -2 \delta^{\dot\beta}_{\dot\alpha}\tcQ^I{}_{\alpha}~.
\label{eq:KQ}
\end{align}
The hermiticity is given by
\begin{align}
&\qquad \qquad (\mathcal{Q}^I{}_\alpha)^\dagger = \mathcal{S}_{I}{}^\alpha~,\qquad (\tilde{\mathcal{Q}}_{I\dot{\alpha}})^\dagger = \tilde{\mathcal{S}}^{I\dot{\alpha}}~,\qquad (\mathcal{R}^I{}_J)^\dagger = \mathcal{R}^J{}_I~,
\nonumber\\
& (\mathcal{M}_{\alpha}{}^\beta)^\dagger = \mathcal{M}_\beta{}^\alpha~,\qquad  (\tilde{\mathcal{M}}_{\alpha}{}^\beta)^\dagger = \tilde{\mathcal{M}}_\beta{}^\alpha~, \qquad (\mathcal{H})^\dagger =\mathcal{H}~,\qquad (\mathcal{P}_{\alpha\dot{\alpha}})^\dagger = \mathcal{K}^{\dot{\alpha}\alpha}~.
\end{align}

\section{Detailed computations}

\label{app:computation2}

We here show that $G\equiv \chi[\mathcal{Q}^3{}_+J^{11}]$  and $\bar{G}\equiv \chi[\tilde{\mathcal{Q}}_{3\dot{+}}J^{11}]$ are Virasoro primaries. To that end, we first recall the argument of sub-section 3.2.4 of \cite{Beem:2013sza}, where the authors proved that any Hall-Littlewood (HL) operators map to Virasoro primaries in two dimensions. Here, HL operators are defined as local operators annihilated by $\cQ^1{}_{-},\,\tcQ_{2\dot\pm}$ and their conjugates, and therefore are Schur operators. The unitarity bounds in \eqref{eq:Higgs_unitarity} imply that HL operators have $E=2R-r,\,j_1 = -r$ and $j_2=0$. Now, suppose that $\{\cO_i\}$ is a basis of the space of Schur operators, and that $\mathcal{O}_1$ is a HL operator. We also use a short-hand notation $\hat{\cO}_i\equiv \chi[\mathcal{O}_i]$ for the corresponding 2d operators. Then the OPE of $\hat{\cO}_1$ and the 2d stress tensor $T(z)$ is of the form
\begin{align}
 T(z)\, \hat{\cO}_1(0) =  \sum_{i}\frac{\hat{\cO}_i(0)}{z^{2+h_1-h_i}}~,
\label{eq:TO}
\end{align}
where $h_i$ is the eigenvalue of $L_0$ for $\hat{\cO}_i$. From equation (3.6) of \cite{Beem:2013sza}, $h_i$ is given by
\begin{align}
h_i = R^{(i)} + j_1^{(i)} + j_2^{(i)},
\end{align}
where $R^{(i)}$ and $(j_1^{(i)},\,j_2^{(i)})$ are the $\SU(2)_R$ charge and the spins of $\mathcal{O}_i$. Since any Schur operator satisfy $r=j_2-j_1$, this is equivalent to
\begin{align}
h_i = R^{(i)} + |r^{(i)}| + 2\,\text{min}\,(j_1^{(i)},j_2^{(i)})~,
\end{align}
where $r^{(i)}$ is  the $U(1)_r$ charge of $O_i$.
Since HL operators have $\text{min}(j_1,j_2)=j_2=0$, we have
\begin{align}
h_1 = R^{(1)} + |r^{(1)}|.
\end{align}
 Therefore \eqref{eq:TO} is rewritten as
\begin{align}
 T(z)\hat{\mathcal{O}}_1(0) =  \sum_{\mathcal{O}_i: \text{ Schur}}\frac{\hat{\mathcal{O}}_i(0)}{z^{2+\Delta R^{(i)}-2\,\text{min}\,(j_1^{(i)},\,j_2^{(i)})}}~,\label{eq:TO2}
\end{align}
where $\Delta R^{(i)}\equiv R^{(1)}-R^{(i)}$. The $\U(1)_R$ charge dependence drops out because $T(z)$ is neutral under $\U(1)_R$.

Recall here that the 2d stress tensor $T(z)$ is given by a linear combination of the 4d $\SU(2)_R$ current $\cJ^{ij}_{+\dot{+}}$ \cite{Beem:2013sza}. Since the $\SU(2)_R$ current is an $\SU(2)_R$ triplet, $T(z)$ is a linear combination of operators with $R=0,\pm 1$. Therefore an OPE with $T(z)$ changes the $SU(2)_R$ charge by $\pm 1$ or $0$, namely
\begin{align}
\Delta R^{(i)} = \pm 1,\; 0~,
\end{align}
depending on $i$. Moreover, from \eqref{eq:Schur}, we see that Schur operators have
\begin{align}
j_1 = \frac{E-2R-r}{2}~,\qquad j_2 = \frac{E-2R+r}{2}~,
\end{align}
whose right-hand sides are positive semi-definite because of the unitarity bounds \eqref{eq:Higgs_unitarity}.
Therefore the worst possible singularity in \eqref{eq:TO2} is of order three. On the other hand, since any Hall-Littlewood operator is a conformal primary, it is annihilated by $L_1$.\footnote{Recall that $L_1$ is identified with $\mathcal{K}^{\dot{+}+}$. See equation (2.19) of \cite{Beem:2013sza}. } Therefore the singularity of order three in \eqref{eq:TO2} vanishes. This means that $\hat{\cO}_1(z)$ is a Virasoro primary.

Now we generalize the above discussion to the cases in which $\cO_1=\cQ^3{}_+J^{11}$ and $\cO_1 = \tcQ_{3\dot{+}}J^{11}$. The only difference is that $\cO_1$ is no longer a HL operator, and therefore \eqref{eq:TO2} is replaced by
\begin{align}
 T(z)\hat{\mathcal{O}}_1(0) =  \sum_{\mathcal{O}_i: \text{ Schur}}\frac{\hat{\mathcal{O}}_i(0)}{z^{\frac{5}{2}+\Delta R^{(i)}-2\,\text{min}\,(j_1^{(i)},\,j_2^{(i)})}}~.
\end{align}
However, the worst possible singularity is still of order three since, as discussed in \cite{Beem:2013sza}, any 2d OPE corresponding to a 4d OPE should be single-valued.  Moreover, since $\mathcal{Q}^3{}_+J^{11}$ and $\tilde{\mathcal{Q}}_{3\dot{+}}J^{11}$ are conformal primaries, the corresponding 2d operator $\hat{\mathcal{O}}_1$ is annihilated by $L_1$. Therefore the worst singularity in the above OPE is of order two. Thus, we see that $G\equiv \chi[\mathcal{Q}^3{}_+J_F^{11}]$  and $\bar{G}\equiv \chi[\tilde{\mathcal{Q}}_{3\dot{+}}J_F^{11}]$ are Virasoro primaries. 

Note here that exactly the same argument tells us that, for any Higgs branch operator $\mathcal{O}$,  it follows that $\chi[\mathcal{O}],\,\chi[\mathcal{Q}^3{}_+\mathcal{O}]$ and $\chi[\tilde{\mathcal{Q}}_{3\dot{+}}\mathcal{O}]$ are Virasoro primaries.\footnote{Note also that the same argument fails in the case of $\hat{\mathcal{O}}_1 = \chi[\frac{1}{2}[\mathcal{Q}^3{}_+,\,\tilde{\mathcal{Q}}_{3\dot{+}}]J_F^{11}]$. In this case, the worst possible singularity in \eqref{eq:TO2} is of order four, which is consistent with the identification of $\chi[\frac{1}{2}[\mathcal{Q}^3{}_+,\,\tilde{\mathcal{Q}}_{3\dot{+}}]J^{11}]$ as the 2d stress tensor.}

\bibliographystyle{ytphys}
\baselineskip=.9\baselineskip
\bibliography{ref}

\end{document}